# Metrics for Computing Trust in a Multi-Agent Environment

Sanat Kumar Bista, Keshav P. Dahal, Peter I. Cowling and Bhadra Man Tuladhar

*Abstract*—One of the risks involved in multi agent community is in the identification of trustworthy agent partners for transaction. In this paper we aim to describe a trust model for measuring trust in the interacting agents. The trust metric model works on the basis of the parameters that we have identified. The model primarily analyses trust value on the basis of the agent's reputation, as provided by the agent itself, and the agent's aggregate rating as provided by the witness agents. The final computation of the trust value is given by a weighted average of these two components. While computing the aggregate rating, a weight based method has been adopted to discount the contribution of possibly un-fair ratings by the witness agents.

## I. INTRODUCTION

Trust is a crucial aspect for any form of interaction. With modern day systems becoming more and more distributed, dynamic and open at the same time, solution approaches involving multi-agent systems, due to its ability to act autonomously and rationally have gained larger interest. The popularity of online trading businesses, virtual organizations over the Internet, the Grid etc, among researchers and its acceptance in the form of widespread use by the users support the fact above. One of the challenges faced by an agent in a virtual community is in assessing the trustworthiness of the other agent with which it wishes to interact. Every agent interacts with others in the community to maximize its own gain, and it can be at the cost of the gain of other agents also [6]. For instance, in an e-commerce setting, where strangers are interacting, it might be more appealing to act deceptively for immediate gain rather than cooperation [2]. As an example, if a buyer pays first, the seller might be tempted to not provide agreed upon quality of goods or services [4]. Thus, it is important for an agent to isolate malicious ones from the reputed and trustworthy. Our approach towards this is to develop a metrics for assessing the worthiness of trust for interacting agents. Our approach of measuring the trust is similar to that of [1]. However, in [1] the authors have primarily focused on the trust metrics in Peer-to-Peer systems. In our case, we consider a general multi-agent environment amenable to trustworthiness assessment problem. This generalization has required us to design a new set of trust parameters. Our model is also capable of handling contributions due to dishonest witness by agents. We further define an approach in combining information from multiple witness agents. The concept of trust network as put forward by the authors in [7], [9] has been explored to provide solutions to the possible dishonest feed back problem.

The major contribution of this paper is in formulating a useful trust metrics model, the outcome of which might be an easily readable figure for interpretation by agents. An enhanced information merging model is presented to bring out new methods in witness analysis and deception detection. In section V of this paper, we present the evaluation and result.

## II. RELATED WORKS

The compelling factor for this research is the absence of a general trust metric system for agent interaction in a multi agent system. There have been researches expressing trust as a probability distribution, which though realistic, is at the same time complex for interpretation [2]. In [1] the authors have proposed a reputation based trust metric system. The metric is mainly targeted towards evaluating trust between peers in a peer-to-peer based electronic community. We have adopted a similar approach to measuring trust, but with enhancements to suit the needs of multi agent systems. In [2], [7] the authors present approaches towards rating aggregations. This involves a series of queries through the network of the agents. In [7] and [9] the authors have defined such a network of agents through which referrals are propagated and computed as *TrustNet*. In computing the aggregate rating for our trust metrics we consider this data structure as a simple referral network. This network is constructed by a requesting agent through referrals by incorporating each referral into the network with *depthLimit* as a bound to the length of chain [9].

Sanat Kumar Bista is with the MOSAIC Research center at the University of Bradford, BD7 1DP, UK as a Research Student. (Corresponding author: phone: ++977-98510-86943; fax: ++977-1-4280612; e-mail: s.k.bista@bradford.ac.uk).

Keshav P. Dahal is with the MOSAIC Research Center at the University of Bradford, BD7 1DP, UK as a Senior Lecturer. (e-mail: k.p.dahal@bradford.ac.uk).

Peter I Cowling is with the MOSAIC Research Center at the University of Bradford, BD7 1DP, UK as a Professor. (e-mail: p.i.cowling@bradford.ac.uk).

Bhadra Man Tuladhar is with the Department of Natural Sciences (Mathematics) at Kathmandu University, Nepal, as a Professor. (e-mail: dean_sci@ku.edu.np).

Approaches towards handling dishonest feedback has also been a focus of research. This is shown by the efforts in [2],[3],[4],[6],[7]and [9]. The authors in these papers have explored multiple ways under Endogenous and Exogenous approaches to the filtering of unfair ratings.

### III. THE SYSTEM

We have identified the following parameters that would contribute towards the trust metric. A listing with brief description is given below:

**1. Agent Reputation**: This is an agent's own version of its reputation acquired by accumulating the positive ratings offered to it by other agents in the past.

**2. Average reputation:** It is the ratio of total number of positive rating score to total number of transactions.

**3. Aggregated Rating:** While considering interaction with unknown agents the subject agent would enquire trust worthiness of the other agent in the agent society to produce an aggregate rating. Calculating such a rating is a non-trivial task. Moreover, the problem is further aggravated by the witness agents providing dishonest feed-backs. Thus, it becomes necessary to compute the aggregate community rating with an inbuilt filtering mechanism for discounting unfair/dishonest ratings from the agent society. Our approach explores the possibility of such a computation.

**4. Agent's Community Guarantee:** An agent might be a member of certain reputed and trustworthy community. Hence, it might sometime also be desirable that the agents own version of its reputation be given a greater weight. This can be adjusted by providing greater weight to the *Agent Community Guarantee* parameter in our model. The *Trust Metric* section below exemplifies a way in which this value can be assigned in the model.

#### A. The Trust Metric

Based on the parameters identified above, we establish a general trust metric formula based on weighted average.

**Establishing the trust metric:** Let X be an agent whose trust value is being measured in any $i^{th}$ instance of time $t$. Let R(X) denote the total reputation possessed by the agent X, which is a direct correlation to the number of successful transactions of X in past, and R(Xi) give the reputation of X at sometime t=i, due to the positive ratings of other agents. Similarly, $n(TR_x)$ gives the total number of transaction (successful as well as unsuccessful) that the agent X has had with all other agents for which it was reputed by them. Let, AGR(X) stand for the aggregate rating of the agent X by all member agents with whom it has interacted in the past. The details of the operating procedure of this function are given in section *B* below.  Also, let CGF(X) represent the community guarantee factor for the agent X. The function CGF (X) represents the contribution of all the communities of which the agent X is a member of. We assume here that an agent if it is not a member of at least one established community is not reliable enough for believing what it says of itself. In other words, in case where CGF(X) has a zero value, the total computation of trust relies on the aggregated rating obtained by the agent from other agents. For simplicity, here we will consider only two states of community guarantee, either an agent is backed up by established community (CGF(X) = 1), or it is not (CGF(X) = 0).

With these, the trust of agent X, T(X) can be defined as:

$$T(x) = WG_a * (CFG(x) * \frac{\sum_{i=1}^{n(TR_x)} R(x_i)}{n(TR_x)}) + WG_b * AGR(x) \quad (1)$$

Where, $WG_a$ and $WG_b$ are the weight factors associated with each of the components of the agent X's trust measurement. Further, $0 \leq WG_a \leq 1$, $0 \leq WG_b \leq 1$, but $WG_a + WG_b = 1$. The full rating scale of trust is $0 \leq T \leq 1$.

#### B. Computing the aggregate rating and Score

In general, each agent is reputed by another agent after each transaction by providing a rating that can either be negative or positive. This rating can be expressed as a vector R of the number of positive (successful) and negative (unsuccessful) ratings. Following [2] we define the vector R and the aggregation process as,

$$R = [S, U] \quad (2)$$

where, *s* stands for successful rating and *u* for unsuccessful rating such that s ≥ 0 and u ≥ 0. One approach of inferring probability of trustworthiness from such data is the use of Beta Probability Distribution functions like in [10]. f(p|α,β) can be expressed using the gamma function Γ as:

$$f(p | \alpha, \beta) = \frac{\Gamma(\alpha + \beta)}{\Gamma(\alpha)\Gamma(\beta)} p^{\alpha-1} (1-p)^{\beta-1} \quad (3)$$

Where $0 \leq p \leq 1, \alpha > 0, \beta > 0$

With a condition that p ≠ 0 if α < 1, and p≠1 if β<1, the probability expectation value of beta distribution is given by:

$$E(p) = \frac{\alpha}{(\alpha + \beta)} \quad (4)$$

Aggregation of ratings is obtained by simple vector addition. Let (X, Z) be a pair of agent. An aggregate rating $R^t$(X, Z) highlighting X's overall opinion of Z at time $t$ is obtained as:

$$R_Z^t(x) = \sum_{t,x} R \quad (5)$$

Using the concepts from equations (3, 4 and 5) a simple point estimate of agent Z's probability distribution after the occurrence of S successful and U unsuccessful events is given by the expected value of the probability distribution as:

$$E(R_w^t(x)) = \frac{(S+1)}{(S+1)+(U+1)} = \frac{S+1}{S+U+2} \quad (6)$$

## IV. REFERRAL NETWORK

A referral network is realized by four different entities, the *requesting agent*, the *target agent*, the *witness agent* and the *references*. First, there is an agent, $A_R$, requesting for reference on some target agent, $A_T$. The agent $A_R$ will query witness agents $W_1$, $W_2$ …$W_n$ on the rating of the target agent $A_T$. The witness agents will provide the feedback based upon each of its experience of working with the target agent in the past. For conformance of the referral results to the aggregation procedure in section (III B) above, we pass the referral result as a rating vector consisting of the number of successful and unsuccessful transactions the witness agent has had with the target agent in the past.

**Filtering Unfair Rating:** While providing the ratings on the target agent by the witness agents, there is every chance that the agent provides an unfair and deceitful feedback. In [2],[3],[4],[7],[9] the authors have given their approaches towards filtering the unfair ratings. Broadly the approaches fall under two different categories [2]:

- **Endogenous discounting of unfair ratings**: In which the statistical properties of the ratings are analyzed to reveal any chances of unfair rating.
- **Exogenous discounting of unfair ratings**: In which the ratings are weighed according to the reputation of the rater.

In our trust metric system we follow the approach in [9], which concentrates on exogenous discounting of unfair ratings. Relying on weighted majority continuous (WMC) (the authors' variation of Weighted Majority Algorithm (WMA)[11]), they compute weight of each witness agent providing the rating, in which the rating expressed as belief function is transformed into probabilities of telling or not telling truth. Thus, the weight $W_i$' of witness $W_i$, with R ($0 \leq R \leq 1$) as the rating from agent $A_R$ is computed as:

$$W_i' = \theta \, W_i \quad (7)$$

Where, $\theta = 1 - ((|prob_i(\{T\}) - R|)/2)$ and, $0 < \theta \leq 1$

Here, $prob_i(\{T\})$ is the probability that the witness agent's feedback on the target agent's *i*th interaction is true and the $prob_i(\{T\})$ is computed by transforming the belief function by applying WMC [9]. The theory that we rely on here is that the witness provided by the agents with greater weight have greater value, and it is normally the agents with lower weight that provide dishonest or unfair rating.

This value of $\theta$, can now be applied as a weight to witness agent. Say, $\theta$ is the weight of witness X for agent Z whose rating is being scored. Thus based on equation (6) Z's reputation score at time *t* is given by:

$$\theta * E(R_w^t(x)) = \theta * \frac{S+1}{S+U+2} \quad (8)$$

The overall computation of AGR(X) in equation (1) can now be obtained by taking an average of the scores weighted by θ.

## V. EVALUATION AND RESULTS

We demonstrate with examples how trust can be computed through our model and what results can be inferred from the values. The context of a calculation like this can be thought of as an electronic commerce environment where a trading agent wishing to transact might consider assessing the trust worthiness of the other agent. The evaluation parameters and relevant assumptions are listed in the table below:

**TABLE I Evaluation Parameters**

| Component | Parameter | Description | Value |
|---|---|---|---|
| Equation (1) first component parameters (agent's own version of data) | $WG_a$ | Weight for first component in equation (1) | 0.5, (*indicating a balanced weight*) |
| | CGF(X) | Community guarantee factor | 1, (*indicating a community guarantee*) |
| | $N(TR_x)$ | Total no. of agent X's transactions | Say, 45 |
| | R(X) | Total reputation of agent X, (*total count of X's successful transactions*) | Say, 25 (*indicating a low success*) |
| Equation (1) second component parameters (witness version of data) | $WG_b$ | Weight for second component in equation (1) | 0.5, (*indicating a balanced weight*) |
| | $RW_1, RW_2, RW_3, RW_4, RW_5$ | Vector representation of witness ratings | Say, [2, 6],[5, 5],[6, 2],[0, 8],[8, 0], Respectively |
| | $\theta_{W1},\theta_{W2},\theta_{W3},\theta_{W4},\theta_{W5}$ | Weight assigned to each witness agent | Say, 0.5, 0.75, 0.8, 0, 1 respectively |

The value for the trust of the agent X, T(X) is obtained after applying the given values to equation (1) and (8). Thus, we get the value for T(X) to be 0.43 for the given values in the table above. Towards this score for T(X), contribution of the first part of equation (1) is 0.28 and that of the second part of equation (1) is 0.15. In real world situations, where a mapping of score to the degree of trustworthiness is defined, a score like what we obtained above for agent X can be readily utilized to decide on whether to partner for business or not. Figure 2 below relates successful/unsuccessful rating to the reputation score. Many of the values assumed above are arbitrarily chosen, however, it has been our attempt to pick values representing diverse conditions. Our system is modular in nature and gives multiple reflections as we modify the parameters. Another interesting part is to compute possible deviation of truth by the agent by comparing the value for the parameters of the first component of equation (1) to the second component (as illustrated in figure 3 below). For instance, the reputation value and the total no. of transaction value in the first component (in which these values are the agent's own version) can be compared to that in the second component (in which these are obtained through social rating). Appropriate measures can be taken if it is found that the corresponding values are significantly different to each other. In our example above, we have the taken the values to be non- conformant to each other, hence there lies a difference in the computed value of first and second part. Some meaningful graphical representations based on the data above are given below:

**TABLE II Some Specific data for analysis**

| Successful Rating | Unsuccessful Rating | Reputation Score (Equation (6)) | Theta | Theta * Reputation Score (Equation (8)) |
|---|---|---|---|---|
| 2 | 6 | 0.3 | 0.5 | 0.15 |
| 5 | 5 | 0.5 | 0.75 | 0.375 |
| 6 | 2 | 0.7 | 0.8 | 0.56 |
| 0 | 8 | 0.1 | 0.01 | 0.001 |
| 8 | 0 | 0.9 | 1 | 0.9 |

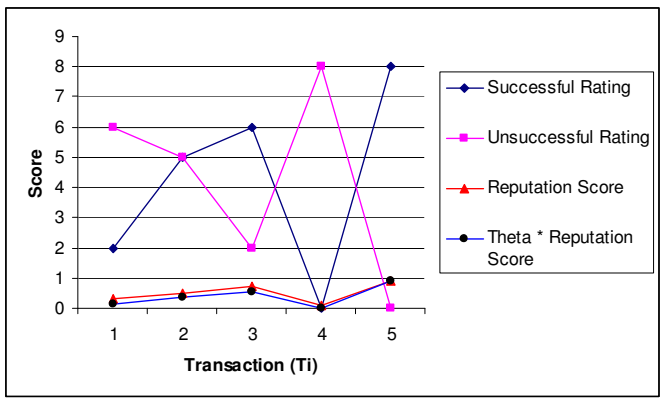

Fig. 2. Comparative view of successful unsuccessful ratings with the reputation score.

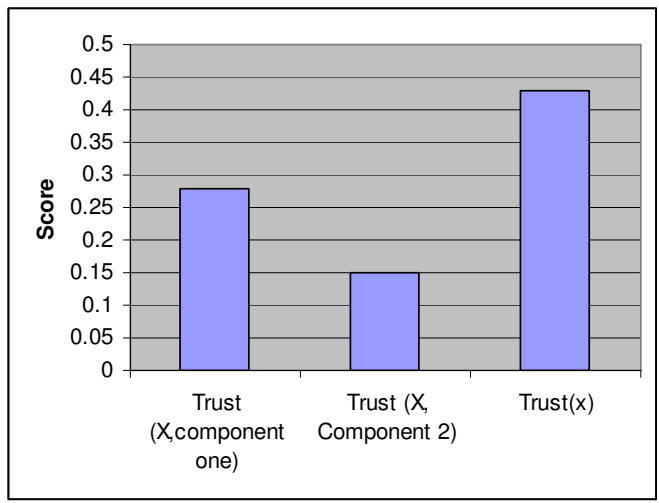

Fig. 3 Component wise contribution of overall trust computation

## VI. DISCUSSION AND CONCLUSION

In this paper we have attempted to establish a trust metric system for a multi agent environment. Our contribution in this has been particularly in deciding the trust parameters, applying them into the trust metrics model, gathering ways to compute aggregate ratings, define a referral network, and apply an approach towards filtering possibly unfair ratings. We are of the opinion that the establishment of the trust parameters and computation of trust value using those parameters have contributed in formulating a simple approach towards trust computation.

A limitation of our model is that it considers only the exogenous approach to filter the possibly unfair ratings. While the application of endogenous approaches could also have been meaningful in filtering the unfair ratings, our model does not currently incorporate this approach. Future work can be in the refinement in terms of finding out ways to merge the Endogenous and Exogenous approach to filter unfair ratings.

## VII. REFERENCES

[1] L. Xiong and L. Liu, "PeerTrust: Supporting Reputation-Based Trust for Peer-to-Peer electronic communities", *IEEE Transactions on Knowledge and Data Engineering,* vol. 16, no. 7, 2004.

[2] A. Whitby, A. JØsang and J. Indulska, "Filtering Out Unfair Ratings in the Bayesian Reputation Systems ", *Icfain Journal of Management Research,* vol. IV, no. 2, pp.48-64, 2005.

[3] C. Dellarocas, "Immunizing online Reputation Reporting Systems Against Unfair Ratings and Discriminatory Behavior", *Proc. Second ACM conf. Electronic Commerce,* 2000.

[4] C. Dellarocas, "The Design of Reliable Trust Management Systems for Electronic Trading Communities", *Working Paper Sloan School of Management MIT,* 2000.

[5] M.Chen and J.P. Singh, "Computing ad Using Reputations for Internet Ratings", *Proc. Third ACM conf. Electronic Commerce,* 2001.

[6] W. T. Teacy, J. Patel, N.R. Jennings and M. Luck, "Coping with Inaccurate Reputation Sources: Experimental Analysis of a Probabilistic Trust Model", *Proc. ACM AAMAS,* 2005.

[7] M. Schillo, P. Funk and M. Rovatsos, "Using Trust for Detecting Deceitful Agents in Artificial Societies", *Applied Artificial Intelligence Journal, Special Issue on Trust, Deception and Fraud in Agent Societies,* 2000.

[8] E.M. Maximilien and M. P. Singh, "Agent-Based Trust Model Involving Multiple Qualities", *Proc. ACM AAMAS,* 2005.

[9] B. Yu and M.P. Singh, "Detecting Deception in Reputation Management", *Proc. ACM AAMAS,* 2003.

[10] Josang, A. and Ismail, R. "The Beta Reputation System". In: *15th Bled Electronic Commerce Conference.* Bled Slovenia. 2002

[11] N. Littlestone and M.K.Warmuth, "The Weighted Majority Algorithm", *Information and Computation,* vol. 2, no. 108, 1994.